\documentclass[intlimits]{amsart}
\usepackage[dvips]{graphicx}

\newcommand{\abs}[1]{\left|#1\right|}           
\newcommand{\mat}[1]{\boldsymbol{#1}}           

\begin{document}
\title{Chaos in a one-dimensional integrable quantum system}
\author{Petr \v Seba${}^{1,2,3}$ and Daniel Va\v sata ${}^{3,4}$}

\maketitle
\begin{center}
\small{${}^1$ University of Hradec Kr\'alov\'e, Hradec
Kr\'alov\'e - Czech Republic}\\
\small{${}^2$ Institute of Physics, Academy of Sciences of the
Czech Republic,\\ Prague - Czech Republic}\\
\small{${}^3$ Doppler Institute for Mathematical Physics and Applied
Mathematics,\\ Faculty of Nuclear Sciences and Physical
Engineering, Czech Technical University, Prague - Czech Republic}\\
\small{${}^4$ Department of Physics, Faculty of Nuclear Sciences and Physical Engineering,\\
Czech Technical University, Prague - Czech Republic}\\
\end{center}

\begin{abstract}
We study a simple one-dimensional quantum system on a circle with
$n$ scale free point interactions. The spectrum of this system is
discrete and expressible as a solution of an explicit secular
equation. However, its statistical properties are nontrivial. The
level spacing distribution between its neighboring odd and even
levels displays a surprising agreement with the prediction obtained
for the Gaussian Orthogonal Ensemble of random matrices.
\end{abstract}
\maketitle
\section{Introduction}
One possible approach to study the chaotic properties of bounded
time-independent quantum system is based on results of the Random
matrix theory (RMT). Bohigas, Giannoni and Schmit \cite{BGS}
conjectured that the local fluctuations of quantum energy levels of
such systems display universal properties. Following this conjecture
the level spacing distribution of integrable systems with more than
one degree of freedom is expected to be Poissonian, while the
distribution for systems that are  classically chaotic is described
by the random matrix theory (see \cite{haake}). (Integrability in
the quantum case means that the eigenvalues and eigenvectors of the
system can be evaluated by solving a simple algebraic equation.)
The specific distribution depends in this case on the symmetry of
the system and is described by the Gaussian orthogonal, unitary or
symplectic ensemble respectively. Nowadays there is  an impressive
amount of evidence for the relevance of this conjecture and so for
the link between the random matrix theory and the quantum behaviour
of classically chaotic systems.

The assumption about the number of degrees of freedom is, however,
quite important since the conjecture doesn't hold in one dimension.
All classical time independent one dimensional systems are
integrable. Nevertheless there are quasi one-dimensional quantum
systems (the quantum graphs) that display random matrix properties.
Recent papers give a wide theoretical and experimental evidence for
such behaviour in various interconnected quantum graphs \cite{sirko,
periodicoth, expspform, qgappltoqchauss}. The quantum dynamic is
here related to the nontrivial topological structure of the graph
and the random matrix description of the spectral statistics of the
quantum graph is based on this fact.

Quantum graphs are topologically nontrivial since each node is
connected with at least three neighbors. This is why these systems
are regarded as quasi one dimensional. In a very one dimensional
system each point is topologically related only to its two
neighbors: to the one on its right and to the one on its left. Our
aim here is to construct a virtually one dimensional system (i.e.
each node has only 2 neighbors) displaying a random matrix behavior.
The work is based on the paper \cite{hejcikcheon}. However there is
a great difference in results. We focus on  the weak coupling limit
and show that in this case, the level-spacing spectral statistic
display two fundamentally different types of behavior. The
statistical properties of the odd level spacings are described by
the random matrix theory  while the even spacings do not show this
behavior.

We deal with a one dimensional particle on a circle with scale free
point interactions localized at the points
$0<x_{1}<\ldots<x_{n}<2\pi$. The scale free point interaction
(\cite{fulop-freepart, fulop-sym-dual}) localized at the origin is
described by a one parameter family of  boundary conditions:
$\frac{1}{\alpha}f(0_-) = f(0_{+}),\quad \alpha f'(0_{-}) =
f'(0_{+}).$ It has the amazing property that the corresponding
reflection / transmission coefficients do not depend on the particle
energy. The scale free point interaction is also expressible as a
limit of short-range potentials, see \cite{chernoff, shigehara}.

The Hamiltonian of the system is represented by the one dimensional
Laplace operator
\begin{equation}
T_{\alpha} = -\frac{d^{2}}{dx^{2}}
\end{equation}
with a  set of boundary conditions:
\begin{eqnarray}
\label{1kruz}f(2\pi_{-})=f(0_{+}),&f'(2\pi_{-}) = f'(0_{+}),\quad\\
\label{2kruz}\frac{1}{\alpha}f(x_{k-}) = f(x_{k+}),&\quad
\alpha f'(x_{k-}) = f'(x_{k+}),\ k=1,\ldots,n,
\end{eqnarray}
where $\alpha\in (0,+\infty)$ is a parameter. We denote the domain
of $T_{\alpha}$ as $D(T_{\alpha})$. The periodic boundary conditions
\eqref{1kruz} define a circle. The second pair \eqref{2kruz}
describes the point interactions. It can be easily shown (see the
theory of self adjoint extensions e.g.\cite{dunford}) that the
operator $T_{\alpha}$ is self adjoint for every $\alpha$. Moreover,
its spectrum is purely discrete containing eigenvalues of finite
multiplicity without limit points. Note that for $\alpha=1$ the
Hamiltonian describes simply a free particle on a circle.
Furthermore in the limits $\alpha \rightarrow 0,\ \alpha \rightarrow
+\infty$ the system decouples into $n$ independent subsystems with
the Dirichlet boundary condition on one side and the Neumann
condition on the other side of the subintervals $(x_k,x_{k+1})$.

\section{The spectrum}
In what follows we will focus on the  positive part of the spectrum
denoting $x_{0} = 0$ and $x_{n+1} = 2\pi$. An eigenvector
$\varphi_{k}, k>0$ corresponding to the energy  $E = k^{2}$ writes
on the subinterval $(x_{j},x_{j+1}),\ j=0,\ldots,n$ as
\begin{equation}
\varphi_{k}(x) = A_j(k)e^{ikx} + B_{j}(k)e^{-ikx},
\end{equation}
with some coefficients $A_j$ and $B_{j}$. The vector
$\varphi_{k}(x)$ belongs to the domain $D(T_{\alpha})$ if it
satisfies the boundary conditions \eqref{1kruz} and \eqref{2kruz}.
This leads to a relation between the neighboring coefficients $A_{j-1},\
B_{j-1}$ and $A_{j},\ B_{j}$
\begin{displaymath}
{A_{j}\choose B_{j}}=\mat{C}_{j}{A_{j-1}\choose B_{j-1}}.
\end{displaymath}
for each $j=1,\ldots,n$, with $C_j$ being a 2x2 matrix:
\begin{equation}\label{tjmat}
\mat{C}_{j} = \frac{1}{2}\left(\begin{array}{cc}\left(\frac{1}{\alpha}+\alpha\right)&
\left(\frac{1}{\alpha}-\alpha\right)e^{-i2kx_{j}}\\
\left(\frac{1}{\alpha}-\alpha\right)e^{i2kx_{j}}&\left(\frac{1}{\alpha}+\alpha\right)\end{array}\right).
\end{equation}
The dependence between $A_n,\ B_{n}$ and $A_{0},\ B_{0}$  is given
by the periodic boundary conditions (\ref{1kruz}) and leads to a
relation
\begin{equation}\label{rel0n}
{A_{0}\choose B_{0}}=\mat{C}_{P}{A_{n}\choose B_{n}}
\end{equation}
with matrix
\begin{equation}\label{cpmat}
\mat{C}_{P} = \left(\begin{array}{cc}e^{i2k\pi}&0\\
0&e^{-i2k\pi}\end{array}\right).
\end{equation}
Combining all these relations together we obtain that the
coefficients $A_{0},\ B_{0}$ determine an eigenvector of
$T_{\alpha}$ iff
\begin{equation}\label{soustkruz}
\left(\mat{C}_{P}\mat{C}_{n}\cdots\mat{C}_{1} - \mat{I}\right){A_{0}\choose B_{0}}=0.
\end{equation}
So the eigenvalues of $T_{\alpha}$ are obtained as solutions of the
equation
\begin{equation}\label{chareqkruz}
\zeta_{P}(k)\equiv\mathrm{det}\left(\mat{C}_{P}\mat{C}_{n}\cdots\mat{C}_{1}
- \mat{I}\right)=0.
\end{equation}

Since the matrices in the above determinant are of size $2 \times
2$, we can easily determine the explicit form of \eqref{chareqkruz}
from \eqref{tjmat} and \eqref{cpmat}  by induction. The secular
equation becomes
\begin{equation}\label{zetap}
\zeta_{P}(k) = \cos(2k\pi) - (1-\beta^{2})^{\frac{n}{2}} +
\beta^{2}\sum_{k<l}^{n}\cos(2k(x_{k}-x_{l}+\pi)) + \beta^{4}\ldots=0
\end{equation}
where
\begin{equation}\label{beta}
\beta = \frac{1-\alpha^{2}}{1+\alpha^{2}}.
\end{equation}
Note that (\ref{zetap}) is a finite series and it ends when the
number of indices in the last sum reaches the highest even number
$m$ with $m<n$. So the last term is
\begin{equation}
\beta^{m}\sum_{i_{1}<\ldots<i_{m}}^{n}\cos(2k(x_{i_{1}}-x_{i_{2}}+\ldots+x_{i_{m-1}}-x_{i_{m}} +\pi)).
\end{equation}

Since the equation \eqref{zetap} contains only even powers of
$\beta$ the positive spectrum of $T_{\alpha}$ is invariant under the
replacement $\alpha \leftrightarrow 1/\alpha$ (it leads to the
change $\beta \leftrightarrow -\beta$).

In the free case ($\alpha=1; \beta = 0$) the equation \eqref{zetap}
simplifies to
\begin{equation}
\zeta_{P}(k) = \cos(2k\pi) - 1=0
\end{equation}
and the solutions are just
\begin{equation}\label{phim}
k_{l} = l, \qquad l\in\mathbb{N},
\end{equation}
with the eigenvalues being doubly degenerate (a consequence of the
rotational symmetry of the system).

\section{Statistical properties of the spectrum}
We start with the free case omitting its non degenerate ground
state. Further we suppose that all the points $0<x_{1}<\ldots<
x_{n}< x_{n+1}=2\pi$ are rationally independent, i.e. we suppose
that the equation
\begin{equation}\label{racnezav}
\sum_{i=1}^{n+1}m_{i} x_{i} = 0,\quad m_{i} \in \mathbb{Z}
\end{equation}
has a solution only for  $m_{i} = 0$ for all $i$.

The states are two times degenerate. Since we want to compare the
fluctuation properties of the energy levels with the random matrix
theory we will not work  directly with the eigenvalues $E_l=k_l^2$
but with the values $e_l=2\vert k_l \vert$. The point is that the
values $e_l$ have not to be unfolded and are of density 1. From now
on whenever we deal with the level spacing statistics we have the
"spectrum" $e_l$ in mind.

The free case is trivial: $e_{2l-1}=e_{2l}=2 l, l=1,2,...$. Defining
the level spacing $s_l=e_{l+1}-e_l$ we immediately see that the
probability density $P(s)$ consists simply of two delta peaks
$P(s)=(\delta(s)+\delta(s-2))/2$. The peak at $0$ comes from the
degeneracy of the eigenvalues.

The question is what happens in the case $\alpha\neq 1$, i.e. when
the point interactions are switched on. The degeneracy vanishes and
the two delta peaks turn into smooth distributions.

To see what happens in that case let us assume a small change of the
parameter $\alpha$ :
$$\alpha = 1 + \delta$$
with $\abs{\delta} \ll 1$. The relation \eqref{beta} leads to
$$\beta = \frac{1-\alpha^{2}}{1+\alpha^{2}} = (-\delta)\frac{2 + \delta}{2+2\delta+\delta^{2}} \approx -\delta.$$

So  the  secular equation \eqref{zetap} can be solved perturbatively
supposing the roots $k_{j}$ in the form
\begin{equation}
k_{j} = j + \beta\lambda_{j},\ j\in \mathrm{N}\cup \{0\}
\end{equation}
with $|\beta\lambda_{j}| \ll 1$.
Substituting it into the equation (\eqref{zetap}) and using the
properties of goniometric functions we get
\begin{eqnarray*}
&&1-\frac{4\pi^{2}\beta^{2}\lambda^{2}_{j}}{2} +\ldots - 1 +\frac{n}{2}\beta^{2} +\\
&&+\beta^{2}\sum_{k<l}^{n}\cos(2j(x_{k}-x_{l}))\left(1 - \frac{4\beta^{2}\lambda^{2}_{j}(x_{k}-x_{l}+\pi)^{2}}{2}+\ldots\right)-\\
&&- \beta^{2}\sum_{k<l}^{n}\sin(2j(x_{k}-x_{l}))\left(2\beta\lambda_{j}(x_{k}-x_{l}+\pi) - \ldots\right) + \beta^{4}\ldots = 0.
\end{eqnarray*}
The absolute term drops out and we get (in the order of $\beta^{2}$)
$$-2\pi^{2}\lambda_{j}^{2} + \frac{n}{2} + \sum_{k<l}^{n}\cos(2j(x_{k}-x_{l})) = 0.$$
The  solutions $\lambda_{j}$ are given by
\begin{equation}\label{lambdapmj}
\lambda^{\pm}_{j} = \pm\frac{1}{2\pi}\sqrt{\left(\sum_{k=1}^{n}\cos(2jx_{k})\right)^{2} +
\left(\sum_{k=1}^{n}\sin(2jx_{k})\right)^{2}}.
\end{equation}

The degenerate eigenvalues split and become
\begin{equation}
k_{2j-1} = j + \abs{\beta}\lambda^{-}_{j}, \quad \quad k_{2j} = j +
\abs{\beta}\lambda^{+}_{j} \quad \quad j=1,2,...
\end{equation}
So the unfolded level spacings are
\begin{equation}\label{spaceexpr}
s_{2j-1} = 4\abs{\beta}\lambda^{+}_{j},\quad s_{2j} = 2 -
2\abs{\beta}(\lambda^{+}_{j+1} + \lambda^{+}_{j}).
\end{equation}
The odd spacings $s_{2j-1}$ contribute to the widening of the
$\delta -$ peak localized at $0$ while the even spacings $s_{2j}$
broaden the $\delta$ peak localized at 2. We will now show that the
peak at 0 is not only broaden but it acquires a shape prescribed by
the random matrix theory.

The positions $x_{i}$  and $\pi$ are altogether rationally
independent. Hence  $x_{i}/\pi$ are also rationally independent
irrational numbers. The periodicity of cosine leads to the relation
$$\cos(2jx_{i}) = \cos(2jx_{i}\ \mathrm{mod}\ 2\pi) = \cos(2\pi((j x_{i}/\pi)\ \mathrm{mod}\ 1)),$$
where mod means the remainder after integer division. From the
theory of distribution of sequences modulo one (see
\cite{sequences}) we know that when $j$ changes the arguments of
cosines and sines in \eqref{lambdapmj} behave like independent
random variables uniformly distributed in $[0,2\pi)$. From this fact
follows that the distribution of the sequence
$\{\cos(2jx_{i})\}_{j=1}^{+\infty}$ has a probability density
function
\begin{equation}\label{distrcos}
f(x)=\left\{\begin{array}{lc}\frac{1}{\pi \sqrt{1-x^{2}}},&x\in(-1,1),\\
0,&x\in\mathbb{R}\setminus(-1,1).\end{array}\right.
\end{equation}
Moreover for different positions $x_i$, $i = 1,\ldots,n$  these
sequences behave like independent random variables. The same holds
for $\{\sin(2jx_{k})\}_{j}$.  So the sums
$\{\sum_{k=1}^{n}\cos(2jx_{k})\}_{j}$,
$\{\sum_{k=1}^{n}\sin(2jx_{k})\}_{j}$ normalized by the factor
$\sqrt{2}/\sqrt{n}$ converge with increasing $n$ to two normally
distributed random variables.

In fact there two  variables are \it statistically independent. \rm
To see this we sketch the proof of this fact. It is based on the
observation that the pair $\left(\cos(X), \sin(X)\right)$, with $X$
being uniformly distributed along $(0,2\pi)$, can be equivalently
represented in the form
$$\left(\mathrm{sgn}(\cos X) |\cos X|, \mathrm{sgn}(\sin X ) \sqrt{1-|\cos X |^2}\right).$$
As one can easily prove, random variables $\mathrm{sgn}(\cos X),
\mathrm{sgn}(\sin X )$ and $|\cos X|$ are stochastically
independent. So the statistical dependence of the cosine and sine is
 fully contained in their absolute values, while their signs are statistically independent.
This observation together with the characteristic function technique
used in the standard proof of the Central limit theorem shows
finally the statistical independence of the above sums.

Summarizing we see that in the weak perturbation limit the
distribution of the odd spacings $s_{2j-1}$ is given by a square
root of a sum of squares of two independent and normally distributed
random variables.  Properly normalized it is nothing but the Wigner
distribution $P_W$ approximating the spacing distribution $P_{GOE}$
of  the Gaussian orthogonal ensemble of random matrices (GOE):
\begin{equation}\label{GOEpdf}
P_W(s) = \frac{\pi}{2} s e^{-\pi s^{2}/4}.
\end{equation}

Remember that during the limiting procedure the perturbation
condition $|\beta\lambda_{j}| \ll 1$ has to be fulfilled for every
$n$. Together with the structure of the formula (\ref{lambdapmj}) it
leads to a condition $|\beta| \ll 1/n$. Since the exact Wigner
distribution is obtained in the limit $n\rightarrow \infty$ the
parameter $\beta$ has to decrease to $0$. It is an open question how
this distribution behaves for increasing $n$ and fixed $\beta$. We
suppose that for fixed  $\beta$ the difference to the Wigner
distribution first decreases with increasing $n$. Than, for some
value of $n$, it reaches a minimum and starts to increase when  $n$
violates the condition $|\beta| \ll 1/n$.

The even spacings $s_{2j}$ behave differently. Substracting the
constant 2 the relation \eqref{spaceexpr} contains a sum of two
variables with a Wigner distribution. So the spacing probability
density behaves as $\approx s^3$ for small $s$ and hence in a way
that is not related to the random matrix theory. The next section
shows that the distribution of the odd spacings follows the random
matrix prediction even in the non perturbative regime.

The topology of the system  (a circle) is of fundamental importance.
To show this we investigate an analogous system on a line segment.
All the parameters of the system remain unchanged. The only
difference is that  the periodic boundary conditions (\ref{1kruz})
are replaced by the Dirichlet ones: $f(0) = f(2\pi)=0$. The spectrum
is again purely discrete and the eigenvalues are determined by the
roots of the secular equation
\begin{eqnarray}
&&\zeta_{L}(k) = \sin(2k\pi)  + \beta\sum_{k}^{n}\sin(2k(x_{k}-\pi)) +\nonumber\\
&&+\beta^{2}\sum_{k<l}^{n}\cos(2k(x_{k}-x_{l}+\pi))+\beta^{3}\sum_{k<l<m}^{n}\ldots=0\label{zetal}
\end{eqnarray}
The free case ($\beta = 0$) leads to unfolded energies $e_{l}=l,
l=1,2,...$ and the probability density $P(s)$ of the level spacings
$s_l=e_{l+1}-e_l$ is just $P(s)=\delta(s-1)$.

For weak coupling one applies the perturbation methods. With the
assumption $k_{j} = j/2 + \beta\gamma_{j}+O(\beta^2),\ j\in
\mathrm{N}$ the result is given by
\begin{equation}
\gamma_{j} = \frac{1}{2 \pi}\sum_{k}^{n}\sin(j x_{k}).
\end{equation}
The distribution of $\gamma_{j}$ tends (after rescaling by the
factor $1/\sqrt{n}$) to a normal distribution with zero mean as $n
\rightarrow \infty$. So the unfolded level spacings, $s_j = 1 + 2
\beta (\gamma_{j+1} - \gamma_{j})$, are in the weak coupling limit
normally distributed and not related to the random matrix theory.

\section{Results}
We solved the equation (\ref{chareqkruz}) finding its first $10^6$
roots for different values of $\alpha$ and $n$. The interaction
positions have been chosen as
\begin{equation}\label{positions}
x_k=\frac{2\pi\sqrt{p_{k}}}{\sqrt{p_{n+1}}}, \quad \quad k=1,2,..n
\end{equation}
with $p_{k}$ denoting the $k-$th prime number.  This choice is
obviously rationally independent and $\eqref{racnezav}$ holds. The
presented results are, however, not influenced by this particular
choice. For different positions satisfying $\eqref{racnezav}$ the
results remain unchanged.

In the perturbative regime the odd spacings follow the Wigner
distribution whereas the even spacings not. Interestingly enough the
spacing statistics behave similarly also beyond the weak coupling
limit.

To compare the obtained level spacing distributions with the results
known from the random matrix theory we have evaluated the difference
functions
\begin{eqnarray}\label{fw}
\Delta F_W &=& \int^{+\infty}_{0} (F(s) - F_{W}(s))^{2}ds\\
\label{fgoe}
\Delta F_{GOE} &=& \int^{+\infty}_{0} (F(s) - F_{GOE}(s))^{2}ds
\end{eqnarray}

where $F_W$ and $F_{GOE}$ is the integrated spacing distributions
for the Wigner  and the exact GOE result respectively:
\begin{equation}\label{wig}
    F_W(s)=\int_0^s P_W(u) du; \quad \quad\  F_{GOE}(s)=\int_0^s P_{GOE}(u) du.
\end{equation}

As an approximation of the exact distribution $P_{GOE}$ we used the
Taylor expansion up to the power 42 for small spacings  and the
Dyson asymptotic result otherwise. This approach is described in
\cite{haake} chapter 4.9. It leads to a difference  between the
Wigner surmise and the  GOE result:
\begin{equation}\label{difergw}
\Delta (F_{GOE} - F_{W}) = \int^{+\infty}_{0}(F_{GOE}(s) - F_{W}(s))^{2}ds = 3.9280 \cdot10^{-5}.
\end{equation}

The  distribution $F(s)$ of the odd level spacings $\lbrace
s_{2j-1}\rbrace^{N}_{1}$ is defined in a usual way,
\begin{equation}
F(s) = \frac{1}{N}\sum_{j=1}^{N}\Theta(s-s_{2j-1}),
\end{equation}
with $\Theta$ being the Heaviside step function and $N$ the number
of spacings taken into account. Its derivative (in the limit
$N\to\infty$) is the level spacing density.

\begin{figure}[ht]
\begin{center}
\includegraphics[width=0.8\textwidth]{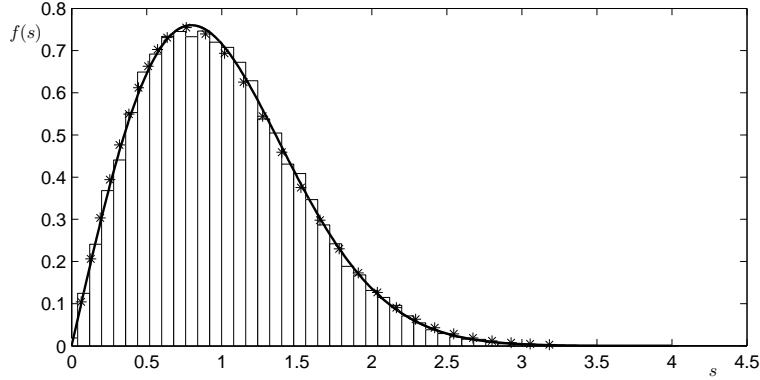}
\end{center}
\caption{The odd level spacing distribution of $s_{2j-1}$ for $47$
point interactions with $\alpha=1.001$ and $N=1\cdot10^{5}$ is
compared with the Wigner distribution (solid line) and the exact GOE
result (stars).} \label{fig1}
\end{figure}

A typical result  is shown on the figure \ref{fig1}. The solid line
marks  the Wigner probability density, \eqref{GOEpdf}. Stars show
the a the level-spacing probability density for GOE as published in
\cite{mehta}, Table A.15.

\begin{figure}[ht]
\begin{center}
\includegraphics[width=0.8\textwidth]{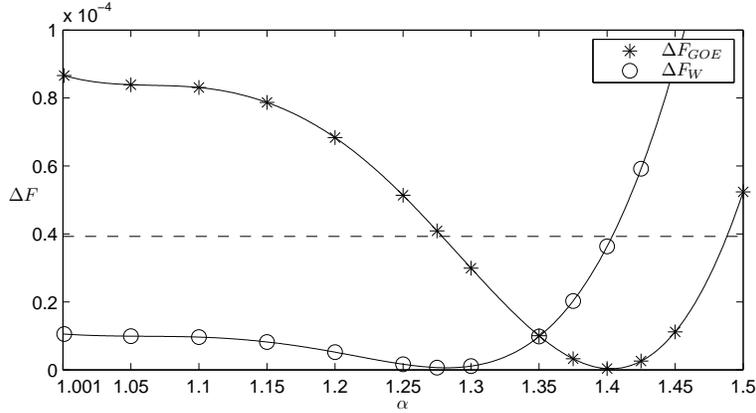}
\end{center}
\caption{Plotted are the differences $\Delta F_{W}$ and $\Delta
F_{GOE}$ for 24 point interactions and $N=10^6$. The dashed
horizontal line represents the difference between true GOE and
Wigner -  \eqref{difergw}.} \label{fig2}
\end{figure}

The dependence of the differences $\Delta F_{W}$ and $\Delta
F_{GOE}$ on the parameter $\alpha$  is shown in the figure
\ref{fig2}

It is not a surprise that for the weak coupling $\alpha \sim 1$ the
sequence $s_{2j-1}$ follows the Wigner distribution \eqref{GOEpdf}
more closely than  Mehta-Gaudin (GOE) distribution. This is just the
consequence of the arguments presented in the previous section. It
remains true, however, also for $\alpha$ up to $\approx 1.35$ i.e.
far outside the weak coupling limit. Even more: there is a distinct
minimum of $\Delta F_{W}$ for $\alpha \approx 1.275$. With
increasing $\alpha$ the distribution starts,however,  approach
closer the GOE result then its Wigner approximation.  For
$\alpha\approx 1.4$ the difference $\Delta F_{GOE}$ displays a clear
minimum with $\Delta_{GOE}=3.6457\cdot 10^{-7}$. For this coupling
the spacing distribution of the model very close to the true GOE
statistic. For $\alpha>1.5$ the agreement however deteriorates step
by step. This is understandable since the system with many
non-periodically placed interactions is influenced by localization
effects for strong enough coupling. For $\alpha\to\infty$ the
spectrum becomes a sum of $n$ mutually embedded equidistant
sequences and becomes Poissonian for rationally independent points
$x_k$ - see \cite{casati}.

\begin{figure}[ht]
\begin{center}
\includegraphics[width=0.8\textwidth]{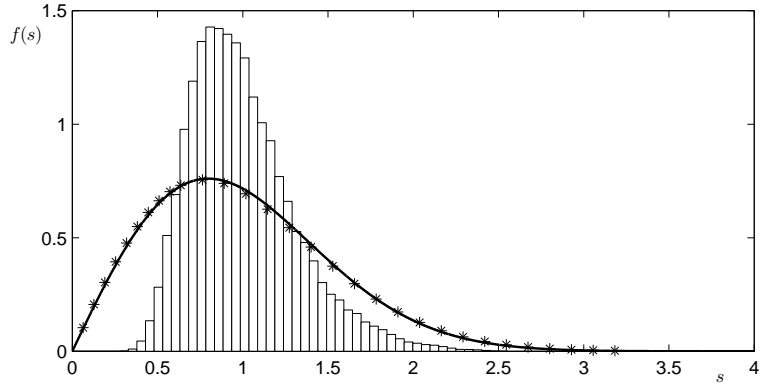}
\end{center}
\caption{The even level spacing probability density for $9$ point
interactions with $\alpha=1.9$ and $N=1\cdot10^{5}$ is compared with
the Wigner and the GOE distributions.} \label{fig3}
\end{figure}

On the other hand the distribution of even spacings $s_{2j}$ never
becomes random matrix like, see figure \ref{fig3}.

\begin{figure}[ht]
\begin{center}
\includegraphics[width=0.8\textwidth]{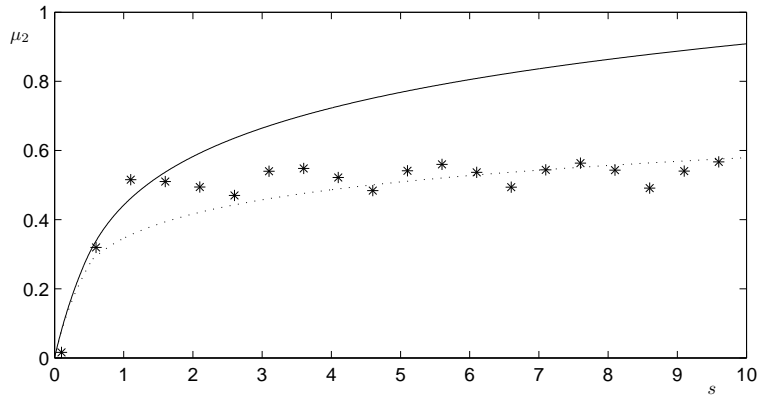}
\end{center}
\caption{The number variance $\mu_2 = \langle n - \langle n \rangle
\rangle^2$ evaluated for 24 point interactions with $\alpha=1.4$ and
$N=1\cdot10^{6}$. The solid and dashed curves correspond  to the
predictions for the  Gaussian orthogonal  (GOE) and  the Gaussian
unitary  (GUE) ensembles respectively.} \label{fig5}
\end{figure}

The level spacing statistic depends only on the neighbouring levels.
More sensitive statistical properties, like for instance the number
variance, describe the correlation of larger level sets. In our
case, such large correlations are dominated by the simple geometric
structure of the model and by the related periodic orbits (see
\cite{berry}). This leads finally to oscillatory pattern like the
one observed in the number variance  and plotted on the figure
\ref{fig5}.

\begin{figure}[ht]
\begin{center}
\includegraphics[width=0.8\textwidth]{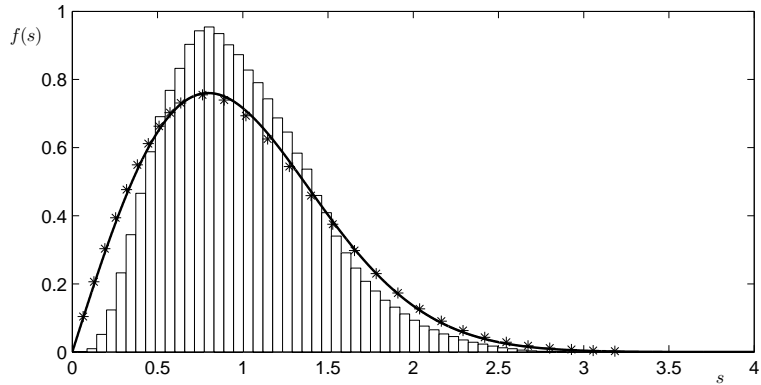}
\end{center}
\caption{The level spacing distribution of $s_{j}$ for $9$ point
interactions on a line segment with $\alpha=1.8$ and
$N=1\cdot10^{5}$ is compared with the Wigner and GOE distributions.}
\label{linefig}
\end{figure}

The spectral properties of the model are closely related to its
topology. Remember that the odd spacings originate from the
splitting of the two times degenerate eigenvalues for the free case
and are in such a way linked to the fact that the system has a
topology of a circle. If we change the topology  and replace the
periodic boundary condition (\ref{1kruz}) with the Dirichlet ones
$f(0) = f(2\pi)=0$ (i.e. we work on a line segment instead of on a
circle) the spacing distribution changes fundamentally and never
follows  the random matrix theory. In this case the distribution of
all spacings (odd and even) is the same. Their common distribution
is displayed on the figure \ref{linefig}.

To summarize: we have discussed a simple one dimensional  quantum
model that displays a level spacing statistics being in a surprising
agreement with the predictions of the Gaussian orthogonal ensemble
of random matrices. We show that this agreement is related to the
topology of the system (circle) and vanishes when the topology is
changed (circle is replaced by a line segment).

{\bf Acknowledgement:} The research was supported by  the Ministry
of Education, Youth and Sports within the project LC06002 and by the
project of the Grant Agency of the Czech Republic No. 202/08/H072.
The stimulating discussions with T.Cheon and P.Hejcik are gratefully
acknowledged. We also thank to the anonymous referee for valuable
remarks.



\begin{thebibliography}{99}
\bibitem{BGS}{Bohigas. O, Giannoni, M. J., Schmit, C.} \emph{Characterization of Chaotic Quantum Spectra and Universality of Level
Fluctuation Laws}, Phys. Rev. Lett. 52 (1984), 1-4

\bibitem{haake}{Haake, F.} \emph{Quantum signatures of chaos}, Springer, Berlin, 1992

\bibitem{sirko}
Hul O., Bauch S., Pakonski P., Savytskyy N., Zyczkowski K., Sirko L.
\emph{Experimental simulation of quantum graphs by microwave
networks}, Phys. Rev. E. 69 (2004), 056205

\bibitem{periodicoth}{Kottos, T., Smilansky, U.} \emph{Periodic Orbit Theory and Spectral Statistics for Quantum Graphs},
Annals of Physics 274 (1998), 76-124

\bibitem{expspform}{Dabaghian, Yu., Bl\"umel, R.} \emph{Explicit Spectral formulas for scaling quantum graphs},
Phys. Rev. E. 70 (2004), 046206


\bibitem{qgappltoqchauss}{Gnutzmann, S., Smilansky, U.} \emph{Quantum Graphs: Application to Quantum Chaos and Universal
Spectral Statistics},  Adv. Phys. 55 (2006), 527-625

\bibitem{fulop-freepart}{F\"ul\"op, T., Tsutsui, I.} \emph{A Free Particle on a Circle with Point Interaction},
Phys. Lett. A 264 (2000), 366-374

\bibitem{fulop-sym-dual}{Cheon, T., F\"ul\"op, T., Tsutsui, I.} \emph{Symmetry, Duality and Anoholonmy of Point
Interaction in One Dimension}, Annals of Physics 294 (2001), 1-23

\bibitem{chernoff}{Chernoff, P.R., Hughes, R.J.} \emph{A New Class of Point Interaction in One Dimension}, J. Funct. Anal. 111 (1993), 97-117

\bibitem{shigehara}{Cheon, T., Shigehara, T.} \emph{Realizing discontinuous wave functions with renormalized short-range potentials}, Phys. Lett. A 243 (1998), 111-116

\bibitem{hejcikcheon}{Hej\v c\'ik, P., Cheon, T.} \emph{Irregular Dynamics in a Solvable One-Dimensional Quantum Graph},
Phys. Lett. A 356 (2006), 290-293

\bibitem{dunford}{Dunford, N., Schwartz, J. T.} \emph{Linear operators - Part II}, New York: John Wiley \& Sons, Inc., 1964

\bibitem{sequences}Kuipers, L., Niederreiter, H. \emph{Uniform Distribution of Sequences},
New York: John Wiley \& Sons, Inc., 1974

\bibitem{mehta}Mehta, M. L. \emph{Random matrices}, Third Edition, San Diego: Elsevier, 2004

\bibitem{casati}Casati G., Chirikov B.V., Guarneri I. \emph{Energy level statistic of integrable quantum systems}
 Phys. Rev. Lett. 54 (1985), 1350 - 1353

\bibitem{berry}Berry, M. V. \emph{Semiclassical theory of spectral rigidity}, Proc. R. Soc. A 400 (1985), 229-251
\end{thebibliography}
\end{document}